\documentstyle[12pt,amssymb]{article}
\def\hybrid{\topmargin 0pt      \oddsidemargin 0pt
        \headheight 0pt \headsep 0pt
        \textwidth 16.5cm
        \textheight 23cm
        \hoffset=0.4cm
        \marginparwidth 0.0in
        \parskip 5pt plus 1pt   \jot = 1.5ex}
\catcode`\@=11
\def\marginnote#1{}

\newcount\hour
\newcount\minute
\newtoks\amorpm
\hour=\time\divide\hour by60
\minute=\time{\multiply\hour by60 \global\advance\minute by-\hour}
\edef\standardtime{{\ifnum\hour<12 \global\amorpm={am}%
        \else\global\amorpm={pm}\advance\hour by-12 \fi
        \ifnum\hour=0 \hour=12 \fi
     \number\hour:\ifnum\minute<10 0\fi\number\minute\the\amorpm}}
\edef\militarytime{\number\hour:\ifnum\minute<10 0\fi\number\minute}

\def\draftlabel#1{{\@bsphack\if@filesw {\let\thepage\relax
   \xdef\@gtempa{\write\@auxout{\string
      \newlabel{#1}{{\@currentlabel}{\thepage}}}}}\@gtempa
   \if@nobreak \ifvmode\nobreak\fi\fi\fi\@esphack}
        \gdef\@eqnlabel{#1}}
\def\@eqnlabel{}
\def\@vacuum{}
\def\draftmarginnote#1{\marginpar{\raggedright\scriptsize\tt#1}}

\def\draft{\oddsidemargin -0.1truein
        \def\@oddfoot{\sl preliminary draft \hfil
        \rm\thepage\hfil\sl\today\quad\militarytime}
        \let\@evenfoot\@oddfoot \overfullrule 3pt
        \let\label=\draftlabel
        \let\marginnote=\draftmarginnote
\def\@eqnnum{{\rm (\theequation)}\rlap
{\kern\marginparsep\tt\@eqnlabel}%
\global\let\@eqnlabel\@vacuum}  }
\newcommand{\R}{{\mathbb{R}}}
\newcommand{\C}{{\mathbb{C}}}
\newfont{\Bbbb}{msbm7 scaled 1\@ptsize00}
\newcommand{\z}{\raise-1pt\hbox{$\mbox{\Bbbb Z}$}}

\font\teneufm=cmmib10 scaled 1\@ptsize00
\font\seveneufm=cmmib7 scaled 1\@ptsize00
\font\fiveeufm=cmmib5  
\def\bfit#1{{\textfont1=\teneufm\scriptfont1=\seveneufm
\scriptscriptfont1=\fiveeufm
\mathchoice{
\hbox{$\mathsurround=0pt\displaystyle#1$}}
{\mathsurround=0pt\hbox{$\textstyle#1$}}
{\hbox{$\mathsurround=0pt\scriptstyle#1$}}
{\hbox{$\mathsurround=0pt\scriptscriptstyle#1$}}}}

\def\numberbysection{\@addtoreset{equation}{section}
        \def\theequation{\thesection.\arabic{equation}}}
\numberbysection

\renewcommand{\theequation}{\thesection.\arabic{equation}}

\def\titlepage{\@restonecolfalse\if@twocolumn\@restonecoltrue\onecolumn
     \else \newpage \fi \thispagestyle{empty}\c@page\z@
        \def\thefootnote{\fnsymbol{footnote}} }

\def\endtitlepage{\if@restonecol\twocolumn \else  \fi
        \def\thefootnote{\arabic{footnote}}
        \setcounter{footnote}{0}}  
\relax

\hybrid

\newtoks\@stequation

\def\subequations{\refstepcounter{equation}%
  \edef\@savedequation{\the\c@equation}%
  \@stequation=\expandafter{\theequation}
  \edef\@savedtheequation{\the\@stequation}
  \edef\oldtheequation{\theequation}%
  \setcounter{equation}{0}%
  \def\theequation{\oldtheequation\alph{equation}}}

\def\endsubequations{%
  \setcounter{equation}{\@savedequation}%
  \@stequation=\expandafter{\@savedtheequation}%
  \edef\theequation{\the\@stequation}%
  \global\@ignoretrue}

\makeatletter
\newdimen\normalarrayskip              
\newdimen\minarrayskip                 
\normalarrayskip\baselineskip
\minarrayskip\jot
\newif\ifold             \oldtrue            \def\new{\oldfalse}
\def\arraymode{\ifold\relax\else\displaystyle\fi}
\def\eqnumphantom{\phantom{(\theequation)}}
\def\@arrayskip{\ifold\baselineskip\z@\lineskip\z@
     \else
     \baselineskip\minarrayskip\lineskip1\baselineskip\fi}

\def\@arrayclassz{\ifcase \@lastchclass \@acolampacol \or
\@ampacol \or \or \or \@addamp \or
   \@acolampacol \or \@firstampfalse \@acol \fi
\edef\@preamble{\@preamble
  \ifcase \@chnum
     \hfil$\relax\arraymode\@sharp$\hfil
     \or $\relax\arraymode\@sharp$\hfil
     \or \hfil$\relax\arraymode\@sharp$\fi}}

\def\@array[#1]#2{\setbox\@arstrutbox=\hbox{\vrule
     height\arraystretch \ht\strutbox
     depth\arraystretch \dp\strutbox
 width\z@}\@mkpream{#2}\edef\@preamble{\halign \noexpand\@halignto
\bgroup \tabskip\z@ \@arstrut \@preamble \tabskip\z@ \cr}%
\let\@startpbox\@@startpbox \let\@endpbox\@@endpbox
  \if #1t\vtop \else \if#1b\vbox \else \vcenter \fi\fi
  \bgroup \let\par\relax
  \let\@sharp##\let\protect\relax
  \@arrayskip\@preamble}
\def\eqnarray{\stepcounter{equation}%
              \let\@currentlabel=\theequation
              \global\@eqnswtrue
              \global\@eqcnt\z@
              \tabskip\@centering             
              \let\\=\@eqncr
              $$%
            \halign to \displaywidth  \bgroup
             \eqnumphantom \@eqnsel
      \hskip\@centering                               
    $\displaystyle  \tabskip\z@ {##}$%
    &\global\@eqcnt\@ne \hskip 2\arraycolsep
         $ \displaystyle  \arraymode{##}$\hfil
    &\global\@eqcnt\tw@ \hskip 2\arraycolsep
         $\displaystyle\tabskip\z@{##}$\hfil
         \tabskip\@centering
    &{##}\tabskip\z@\cr}
\makeatother
\newtheorem{th}{Theorem}[section]  

\newtheorem{rem}{Remark}[section]
\def\bea{\begin{eqnarray}}
\def\eea{\end{eqnarray}}

\def\beq{\begin{equation}}
\def\eeq{\end{equation}}
\def\be{\beq\new\begin{array}{c}}  
\def\ee{\end{array}\eeq}           
\def\bse{\begin{subequations}}                
\def\ese{\end{subequations}}                 %

\def\square{\hfill{\vrule height6pt width6pt            
depth1pt} \break \vspace{.01cm}}                        

\def\bgamma{{\bfit\gamma}}

\def\blambda{{\bfit\lambda}}
\def\brho{{\bfit\rho}}

\def\la{\lambda}

\def\e{\epsilon}
\def\<{\langle}
\def\>{\rangle}

\def\N{\scriptscriptstyle N}

\begin{document}
\begin{titlepage}

\begin{center}

\phantom.
\bigskip\bigskip\bigskip\bigskip\bigskip\bigskip
{\Large\bf Eigenfunctions of $GL(N,\R)$ Toda chain:\\

\bigskip
The Mellin-Barnes representation.}\\

\bigskip \bigskip
{\large S. Kharchev\footnote{E-mail:  kharchev@vitep5.itep.ru},
D. Lebedev\footnote{E-mail:  dlebedev@vitep5.itep.ru}}\\ \medskip
{\it Institute of Theoretical \& Experimental Physics\\ 117259
Moscow, Russia}\\
\end{center}

\vspace{4cm}

\begin{abstract}
\noindent
The recurrent relations between the eigenfunctions for
$GL(N,\R)$ and $GL(N-1,\R)$ quantum Toda chains is derived.
As a corollary, the Mellin-Barnes integral representation for the
eigenfunctions of a quantum open Toda chain is constructed for the
$N$-particle case.
\end{abstract}

\end{titlepage}
\clearpage \newpage

\setcounter{page}1
\footnotesize
\normalsize

\section{Introduction}
Recently a new method of construction of the eigenfunctions for
the periodic Toda chain have been introduced \cite{KHL}. It relies
on generalized Fourier transform expansion over eigenfunctions of
an open Toda chain which coincide with the Whittaker functions of
the $GL(N\!-\!1,\R)$ group
\cite{Jac}-\cite{Ha}.

\noindent
In the original papers \cite{Jac}-\cite{Ha} the Whittaker
functions
were constructed by purely algebraic methods
in terms of the Iwasawa
decomposition for the corresponding group. But it turns out that
there is an alternative way to construct these functions directly
on the level of $R$-matrix formalism using essentially the
integrable properties of the model.

\noindent
This paper was inspired by one remark of E. Sklyanin,
that integral
formula for the eigenfunction of periodic Toda chain \cite{KHL}
can be used to obtain some recurrent relations for $N$-particles
eigenfunctions
of open Toda chain through the $N-1$ particles ones. Sklyanin's
motivations based on a possibility to introduce a formal parameter
to the Baxter equation \cite{Skl}
and then, putting it equal to zero, to transform the second order
difference equation to the first order one.
In our paper we find rigorous proof of this observation
which is free of the limiting procedure and
find an universal way to construct
"auxiliary" eigenfunctions staying completely in the framework of
the $R$-matrix formalism without any reference to an algebraic
scheme developed in \cite{Jac}-\cite{Ha}.
To compare with the well known results on eigenfunctions of open
Toda chain \cite{Konst,St}, our approach presents not only a new
integral representation for the eigenfunctions but, together with
generalized Fourier transformation, it gives a self-consistent
method to solve the spectral problem for periodic Toda chain.
We hope that this method can be applied for some other classes of
integrable systems.

\section{The model}
We start with the $R$-matrix formalism for the quantum periodic
Toda chain \cite{Skl}. Let
\be\label{m1}
L_n(\la)\,=\, \left(\begin{array}{cc}\la-p_n & e^{-x_n}\\ -e^{x_n}
& 0
\end{array}\right)
\ee
be the corresponding Lax operator where
$[x_n,p_m]=i\hbar\delta_{nm}$.
The $N$-particle monodromy matrix
\be\label{m2}
T_{\N}(\la)\;\stackrel{\mbox{\tiny def}}{=}\;L_{\N}(\la)\ldots
L_1(\la)\,\equiv\, \left(\begin{array}{cc}A_{\N}(\la) &
B_{\N}(\la)\\
C_{\N}(\la) & D_{\N}(\la)\end{array}\right)
\ee
satisfies the standard $RTT$ relations with the rational
$R$-matrix. In particular,
\be\label{m3}
(\la-\mu+i\hbar)A_{\N}(\mu)C_{\N}(\la)\,=\,
(\la-\mu)C_{\N}(\la)A_{\N}(\mu)+i\hbar A_{\N}(\la)C_{\N}(\mu)
\ee

\bigskip\noindent
The eigenfunctions for the periodic spectral problem have been
constructed \cite{KHL} with the help of Weyl invariant function
$\psi_{\gamma_1,\ldots,\gamma_{N-1}}(x_1,\ldots,x_{{\N}-1})\equiv
\psi_{\bgamma}(\bfit x)$ which is fast decreasing in the regions
$x_k\gg x_{k+1},\;(k=1,\ldots, N-1)$ and satisfies to equations
\be\label{m4}
C_{\N}(\la)\psi_\bgamma =-\,e^{x_N}\prod_{m=1}^{N-1}
(\la\!-\!\gamma_m)\, \psi_\bgamma
\ee
\be\label{m5}
\hspace{5cm}
A_{\N}(\gamma_j)\psi_\bgamma=i^{-N}e^{-x_N}
\psi_{\bgamma-i\hbar\bfit e_j}
\hspace{1.5cm}(j=1,\ldots, N-1)
\ee
where $\bfit e_j$ is $j$-th basis vector in $\R^{N-1}$.
It is easy to see that the equation (\ref{m5}) is compatible with
(\ref{m4}) due to commutation relation (\ref{m3}).

\bigskip\noindent
Actually, the function $\psi_\bgamma(\bfit x)$ is an appropriate
solution for $N-1$ particle open Toda chain. Indeed,
the operator $A_{{\N}-1}(\la)$
(arising from the $N\!-\!1$ particle problem) is the generating
function for the Hamiltonians of the $GL(N-1,\R)$ Toda chain.
Using the obvious relations between the elements of the monodromy
matrices $T_{\N}(\la)$ and $T_{{\N}-1}(\la)$:
\be\label{m6}
A_{\N}(\la)=(\la\!-\!p_{\N})A_{{\N}-1}(\la)+
e^{-x_N}C_{{\N}-1}(\la)\\
C_{\N}(\la)=-e^{x_N}A_{{\N}-1}(\la)
\ee
the equations (\ref{m4}) and (\ref{m5})
can be written in the equivalent form
\be\label{m7}
A_{{\N}-1}(\la)\psi_\bgamma=\prod_{m=1}^{N-1}(\la\!-\!\gamma_m)
\,\psi_\bgamma
\ee
\be\label{m8}
\hspace{6cm}
C_{{\N}-1}(\gamma_j)\psi_\bgamma=i^{-N}
\psi_{\bgamma-i\hbar\bfit e_j}
\hspace{1.5cm}(j=1,\ldots, N-1)
\ee
These equations fix (up to $i\hbar$-periodic
common factor) the Weyl invariant Whittaker function for
$GL(N\!-\!1,\R)$ group. In \cite{KHL} we choose the factor in
such a way that $\psi_\bgamma$ is an entire function in
$\bgamma$ and the following asymptotics hold:
\be\label{as2}
\psi_\bgamma\,\sim\,|\gamma_j|^{\frac{2-N}{2}}
\exp\Big\{-\frac{\pi}{2\hbar}(N\!-\!2)|\gamma_j|\Big\}
\ee
as $|{\rm Re}\,\gamma_j|\to\infty$ in the finite strip of complex
plane.
\begin{rem}
The function $C_{{\N}-1}(\la)\psi_\bgamma$ is a
polynomial in $\la$ of order $N-2$. Therefore, this polynomial is
restored by their $N-1$ values at given points $\gamma_1,\ldots,
\gamma_{{\N}-1}$.
Hence, one obtains the interpolation formula
\be\label{m9}
C_{{\N}-1}(\la)\psi_\bgamma=i^{-N}\,\sum_{j=1}^{N-1}
\psi_{\bgamma-i\hbar\bfit e_j}\,\prod_{m\neq j}
\frac{\la\!-\!\gamma_m}{\gamma_j\!-\!\gamma_m}
\ee
\end{rem}
Let us introduce the key object - the {\it auxiliary} function
\be\label{m10}
\Psi_{\bfit\gamma,\e}(x_1,\ldots, x_{\N})
\;\stackrel{\mbox{\tiny def}}{=}
\;e^{\raise2pt\hbox{$\frac{i}{\hbar}
\scriptstyle\big(\e\,-\!\sum\limits_{m=1}^{N-1}\gamma_m\big)$}
\raise2pt\hbox{$\scriptstyle x_{\N} $}}\,\psi_{\bfit\gamma}
(\bfit x)
\ee
where $\e$ is an arbitrary parameter.
From (\ref{m7}), (\ref{m9}), and (\ref{m6}) it is readily seen that
this function satisfies to equations
\be\label{m11}
A_{\N}(\la)\Psi_{\bgamma,\e}=
\Big(\la-\e+\sum_{m=1}^{N-1}\gamma_m\Big)
\prod_{j=1}^{N-1}(\la\!-\!\gamma_j)\;\Psi_{\bgamma,\e}+
i^{-N}\sum_{j=1}^{N-1}\Psi_{\bfit\gamma-i\hbar\bfit e_j,\e}\,
\prod_{m\neq j}\frac{\la\!-\!\gamma_m}{\gamma_j\!-\!\gamma_m}
\ee
\be\label{m12}
C_{\N}(\la)\Psi_{\bgamma,\e}=-\,e^{x_N}\prod_{j=1}^{N-1}
(\la-\gamma_j)
\;\Psi_{\bgamma,\e}
\ee

\section{The problem}
Let the Weyl invariant Whittaker function
$\psi_{\gamma_1,\ldots,\gamma_{{\N}-1}}(x_1,\ldots,x_{{\N}-1})$
for $GL(N\!-\!1,\R)$ Toda chain is given.
The problem is to find the corresponding solution
for $GL(N,\R)$ Toda chain using the above information, i.e.
to construct the Weyl invariant Whittaker function
$
\psi_{\la_1,\ldots,\la_{\N}}(x_1,\ldots,x_{\N})
$
satisfying to equations
\bse\label{p1}
\be\label{p1a}
A_{\N}(\la)\psi_{\la_1,\ldots,\la_N}=
\prod_{k=1}^N(\la-\la_k)\;\psi_{\la_1,\ldots,\la_N}
\ee
\be\label{p1b}
\hspace{4.5cm}
C_{\N}(\la_n)\psi_{\la_1,\ldots,\la_N}=
i^{-N-1}\,\psi_{\la_1,\ldots,\la_n-i\hbar,\ldots,\la_N}
\hspace{1cm}(n=1,\ldots, N)
\ee
\ese
and obeying the asymptotics
\be\label{as3}
\psi_{\la_1,\ldots,\la_{\N}}\,\sim\,|\la_n|^{\frac{1-N}{2}}
\exp\Big\{-\frac{\pi}{2\hbar}(N\!-\!1)|\la_n|\Big\}
\ee
as $|{\rm Re}\,\la_n|\to\infty$, in terms of the function
$\psi_\bgamma(\bfit x)$. It is clear that the action of the
operators
$A{\N}(\la)$ and $C_{\N}(\la)$ is nicely defined when acting on the
auxiliary functions $\Psi_{\bgamma,\e}(x_1,\ldots,x_{\N})$.
Therefore, it is reasonable to assume that the solution for
$GL(N,\R)$ Toda chain is described by appropriate (generalized)
Fourier transformation of the function
$\Psi_{\bgamma,\e}(x_1,\ldots,x_{\N})$. This is in complete
analogy with the corresponding construction for the periodic
case \cite{KHL}.

\section{Main statements}
\begin{th}
Let $\Psi_{\bgamma,\e}(\bfit x,x_{\N})$ be the auxiliary function
for $N$-periodic Toda chain, i.e. it is defined in terms of
the Weyl invariant Whittaker function for $GL(N\!-\!1,\R)$ Toda
chain according to (\ref{m10}). Let
$\blambda=(\la_1,\ldots,\la_{{\N}})\in\C^N$ be the set of
indeterminates. Let
\be\label{re2'}
\mu(\bgamma)=(2\pi\hbar)^{N-1}(N\!-\!1)!\prod_{j<k}
\left|\Gamma\Big(\frac{\gamma_j\!-\!\gamma_k}{i\hbar}\Big)\right|^2
\ee
\be\label{re2}
Q(\gamma_1,\ldots,\gamma_{{\N}-1}|\lambda_1,\ldots,\lambda_{\N})=
\prod_{j=1}^{N-1}\prod_{k=1}^N
h^{\frac{\scriptstyle\gamma_j-\la_k}{\scriptstyle i\hbar}}\,
\Gamma\Big(\frac{\gamma_j\!-\!\la_k}{i\hbar}\Big)
\ee
Then the Weyl invariant Whittaker function for $GL(N,\R)$ Toda
chain is given by recurrent formula
\be\label{re4}
\psi_{\la_1,\ldots,\la_N}(x_1,\ldots,x_{\N})=
\int\limits_{\cal C}
\mu^{-1}(\bgamma)Q(\bgamma;\blambda)\Psi_{\bfit\gamma;
\la_1+\ldots+\la_N}(x_1,\ldots,x_{\N})\,d\bgamma
\ee
where the integration is performed along the horizontal lines
with $\mbox{\rm Im}\,\gamma_j>
\mbox{\rm max}_k\,\{\mbox{\rm Im}\,\la_k\}$.
\end{th}

\section{Proof of the Theorem}
First of all, the integral (\ref{re4}) is correctly defined.
Indeed, the function
\be
q(\gamma|\lambda_1,\ldots,\lambda_{\N})\equiv
\prod_{k=1}^Nh^{\frac{\scriptstyle\gamma-\la_k}
{\scriptstyle i\hbar}}
\,\Gamma\Big(\frac{\gamma\!-\!\la_k}{i\hbar}\Big)
\ee
obeys the asymptotics
\be\label{as4}
q(\gamma;\blambda)\sim |\gamma|^{-N/2}
\exp\Big\{-\frac{\pi N}{2\hbar}\,|\gamma|\Big\}
\ee
as $|{\rm Re}\,\gamma|\to\infty$ in finite horizontal strip while
\be
\mu^{-1}(\bgamma)\sim |\gamma_j|^{N-2}
\exp\Big\{\frac{\pi}{\hbar}(N-2)\,|\gamma_j|\Big\}
\ee
Hence,
the integral in (\ref{re4}) is absolutely convergent due to
asymptotics (\ref{as2}).

\bigskip\noindent
Let us verify that the relation (\ref{p1a})
holds. Using (\ref{m11}) one finds
\be\label{xx}
A_{\N}(\la)\psi_{\la_1,\ldots,\la_{\N}}(x_1,\ldots,x_{\N})=\\ =
\int\limits_{\cal C}\Big(\la\!-\!\sum_{k=1}^N\!\la_k\!+
\!\sum_{m=1}^{N-1}\!\gamma_m\Big)
\prod_{j=1}^{N-1}(\la\!-\!\gamma_j)\,
\mu^{-1}(\bgamma)Q(\bgamma;\blambda)\,
\Psi_{\bgamma,\la_1+\ldots+\la_N}(x_1,\ldots,x_{\N})\,d\bgamma\;+\\
+\;i^{-N}\sum_{j=1}^{N-1}\int\limits_{\cal C}
\prod_{m\neq j}\frac{\la\!-\!\gamma_m}{\gamma_j\!-\!\gamma_m}\;
\mu^{-1}(\bgamma)Q(\bgamma;\blambda)\,
\Psi_{\bgamma-i\hbar\bfit e_j,\la_1+\ldots+\la_N}
(x_1,\ldots,x_{\N})\;d\bgamma
\ee
We shift the appropriate integrations $\gamma_j\to\gamma_j+i\hbar$
and use the functional equation
\be\label{pr5}
\mu^{-1}(\bfit\gamma+i\hbar\bfit\delta_j)\,=\,(-1)^N\mu^{-1}
(\bfit\gamma)\,
\prod_{m\neq j}\frac{\gamma_j-\gamma_m\!+\!i\hbar}
{\gamma_j-\gamma_m}
\ee
The second integrand in (\ref{xx}) has no poles in any finite
horizontal strip in the upper half-plane ${\rm Im}\,\gamma_j\!>
\!{\rm max}_k\{{\rm Im}\,\la_k\}$ (all possible poles are
cancelled by appropriate zeros of the function
$\mu^{-1}(\bgamma)$).
Moreover, the intergand is fast decreasing in this strip as
${\rm Re}\,\gamma_j\to\pm\infty$. As a consequence, the integral
over the strip vanishes. Therefore, it is possible to deform the
shifted contour to the original one. Hence, one arrives at
the relation
\be\label{pr6}
A_{\N}(\la)\psi_{\la_1,\ldots,\la_{\N}}= \int\limits_{\cal C}
\!\left\{\!\Big(\la\!-\!\sum_{k=1}^N\!\la_k\!+\!
\sum_{m=1}^{N-1}\!\gamma_m\Big)
\prod_{j=1}^{N-1}(\la\!-\!\gamma_j)
Q(\bgamma;\blambda)\,+\right.\\
+\left.i^N\sum_{j=1}^{N-1}
\prod_{m\neq j}\frac{\la\!-\!\gamma_m}{\gamma_j\!-\!\gamma_m}
\,Q(\bgamma\!+\!i\hbar\bfit e_j;\blambda)\!\right\}\!
\mu^{-1}(\bgamma)
\Psi_{\bgamma,\la_1+\ldots+\la_N}d\bgamma
\ee
The function $Q(\bgamma;\blambda)$ satisfies to equations
\be\label{pr7}
\prod_{k=1}^N(\gamma_j-\la_k)\;Q(\bgamma;\blambda)=i^N
Q(\bgamma+i\hbar\bfit e_j;\blambda)
\ee
for any $j=1,\ldots,N\!-\!1$. Therefore, the relation
(\ref{pr6}) acquires the form
\be\label{pr8}
A(\la)\psi_{\la_1,\ldots,\la_{\N}}=
\int\limits_{\cal C}\!
\left\{\Big(\la\!-\!\sum_{k=1}^N\la_k\!+\!
\sum_{m=1}^{N-1}\!\gamma_s\Big)
\prod_{j=1}^{N-1}(\la\!-\!\gamma_j)\,+\right.\\ +\left.\,
\sum_{j=1}^{N-1}\prod_{k=1}^N(\gamma_j-\la_k)\,
\prod_{m\neq j}\frac{\la\!-\!\gamma_m}{\gamma_j\!-\!\gamma_m}\;
\right\}\mu^{-1}(\bgamma)Q(\bgamma;\blambda)
\Psi_{\bgamma,\la_1+\ldots+\la_N}d\bgamma
\ee
But the expression in curly brackets is nothing but the polynomial
$\prod\limits_{k=1}^N(\la-\la_k)$.
Indeed, any polynomial with the leading terms
$F(\la)=\la^N+f_1\la^{N-1}+\ldots$ can be uniquely restored by its
values at any $N\!-\!1$ arbitrary points
$\gamma_1,\ldots,\gamma_{{\N}-1}$ according
to interpolation formula
\be\label{pr9}
F(\la)=\Big(\la+f_1+\sum_{m=1}^{N-1}\gamma_m\Big)
\prod_{j=1}^{N-1}(\la-\gamma_j)+\sum_{j=1}^{N-1}F(\gamma_j)\,
\prod_{m\neq j}\frac{\la\!-\!\gamma_m}{\gamma_j\!-\!\gamma_m}
\ee
- in our case
\be
F(\la)=\prod\limits_{k=1}^N(\la-\la_k)
\ee
with $f_1=-\la_1-\ldots-\la_{{\N}}$. Hence, we obtain (\ref{p1a}).

\bigskip\noindent
Further, we consider the relation (\ref{p1b}). Using (\ref{m12})
one obtains
\be\label{c1}
C_{\N}(\la_n)\psi_{\la_1,\ldots,\la_{\N}}=
-e^{x_N}\int\limits_{\cal C}
\mu^{-1}(\bgamma)Q(\bgamma;\blambda)\prod_{j=1}^{N-1}
(\la_n-\gamma_j)
\Psi_{\bgamma,\la_1+\ldots+\la_{\N}}\,d\bgamma
\ee
Clearly,
\be
e^{x_N}\Psi_{\bgamma,\la_1+\ldots+\la_{\N}}=
\Psi_{\bgamma,\la_1+\ldots+\la_{\N}-i\hbar}
\ee
and, therefore, the relation (\ref{c1}) acquires the form
\be\label{c2}
C_{\N}(\la_n)\psi_{\la_1,\ldots,\la_{\N}}=
(-1)^N\int\limits_{\cal C}
\mu^{-1}(\bgamma)Q(\bgamma;\blambda)
\prod_{j=1}^{N-1}(\gamma_j-\la_n)
\Psi_{\bgamma,\la_1+\ldots+\la_{\N}-i\hbar}\,d\bgamma
\ee
Evidently, the function $Q(\bgamma;\blambda)$ satisfies to equation
\be\label{pr10}
\prod_{j=1}^{N-1}(\gamma_j-\la_n)Q(\bgamma;\blambda)=
i^{N-1}Q(\bgamma;\blambda-i\hbar \bfit e_n)
\ee
Hence, we prove that function (\ref{re4}) obeys the relations
(\ref{p1b}).

\noindent
The final step is to prove that the function (\ref{re4})
is the genuine Whittaker function.
The integrand in (\ref{re4}) decreases exponentially as
$\gamma_j\to-i\infty\,,\,(j=1,\ldots,N\!-\!1)$ and, as consequence,
the integrals over large semi-circles in the lower half-plane
vanish. Using the Cauchy formula to calculate the integral
(\ref{re4}) in the asymptotic region
$x_{k+1}\gg x_k,\;(k=1,\ldots,N\!-\!1)$,
it is easy to see that the asymptotics of the function
$\psi_{\la_1,\ldots,\la_N}$ are determined precisely in terms of
the corresponding Harish-Chandra functions (see, for example,
\cite{Ha}):
\be\label{exp2}
\psi_\blambda(\bfit x)=
\sum_{s\in W}\hbar^{-2i(s\blambda,\brho)/\hbar}
\prod_{j<k}\Gamma\Big(\frac{s\la_j\!-\!s\la_k}{i\hbar}\Big)
e^{\frac{i}{\hbar}(s\blambda,\bfit x)}+
O\Big(\mbox{\rm max}\Big\{e^{x_k-x_{k+1}}\Big\}_{k=1}^{\N-1}\Big)
\ee
(in the last formula the summation is performed
over the Weyl group).
Hence, we construct exactly the Weyl invariant Whittaker function.
Moreover, using the Stirling formula for the $\Gamma$-functions,
it is easy to see that the asymptotics (\ref{as3}) hold. Theorem is
proved.
\square

\section{The Mellin-Barnes representation}
\begin{th}
Let a set $||\gamma_{jk}||$ be the lower triangular
$N\times N$ matrix. The solution to eqs.(\ref{p1}) can be written
(up to inessential numerical factor) in the form of multiple
Mellin-Barnes integrals:
\be\label{mb}
\psi_{\gamma_{_{{\scriptstyle{\N}}1}},
\ldots,\gamma_{_{\scriptstyle\N \N}}}
(x_1,\ldots ,x_N)\;=\\ =
\int\limits_{\cal C} \prod_{n=1}^{N-1}
\frac{\prod_{j=1}^n\prod_{k=1}^{n+1}
\hbar^{\frac{\scriptstyle\gamma_{nj}-\gamma_{n+1,k}\!}
{\scriptstyle i\hbar}}\,
\Gamma\Big(\frac{\textstyle\gamma_{nj}
\!-\!\gamma_{n+1,k}}{\textstyle i\hbar}\Big)}
{\prod\limits_{\stackrel{\scriptstyle j,k=1}{j<k}}^n
\Big|\Gamma\Big(\frac{\textstyle\gamma_{nj}-\gamma_{nk}}
{\textstyle i\hbar}\Big)\Big|^2}\,
\exp\left\{\frac{i}{\hbar}\sum_{n,k=1}^{N}x_n
\Big(\gamma_{nk}-\gamma_{n-1,k}\Big)\right\}
\prod\limits_{\stackrel{\scriptstyle j,k=1}{j\leq k}}^{N-1}
d\gamma_{jk}
\ee
where the integral should be understand as follows:
first we integrate on  $\gamma_{11}$ over the line
${\rm Im}\,\gamma_{11} >
\max\{{\rm Im}\,\gamma_{21}, {\rm Im}\,\gamma_{22}\}$;
then we integrate on the set $(\gamma_{21} ,\gamma_{22})$ over
the lines ${\rm Im}\,\gamma_{2j} >
\max_{m}\{{\rm\,Im}\gamma_ {3m}\}$ and so on. The last integrations
should be performed on the set of variables
$(\gamma_{{\N}-1,1}\ldots,\gamma_{{\N}-1,{\N}-1})$
over the lines
${\rm Im}\gamma_{{\N}-1,k}>\max_{m}\{{\rm Im}\,\gamma_{{\N},m}\}$.
\end{th}
The proof is straightforward resolution of the recurrent relations
(\ref{re4}) starting with trivial Whittaker function
$\psi_{\gamma_{_{11}}}(x_1)=\exp\{\frac{i}{\hbar}\gamma_{11} x_1\}$.

\section*{Acknowledgments}

We are deeply indebted to E.Sklyanin for drawing our attention on
the possibility of recurrent reconstruction of the eigenfunction of
the open Toda chain. We thanks to M.Semenov-Tian-Shansky
for stimulating discussions.

\bigskip\noindent
D.Lebedev would like to thank F.Smirnov and
L.P.T.H.E., Universit\'e Pierre \'et Marie Curie for hospitality,
where the work was partially done.
The research was partly supported by grants INTAS 97-1312; RFFI
98-01-00344 (S. Kharchev); RFFI 98-0100328 (D.Lebedev) and by grant
96-15-96455 for Support of Scientific Schools.


\begin{thebibliography}{12}

\bibitem{KHL}S.Kharchev, D.Lebedev, {\it Integral representation
for the eigenfunctions of quantum periodic Toda chain},
hep-th/9910265 (to be publ. in Lett.Math.Phys.)

\bibitem{Jac}H.Jacquet, {\it Fonctions de Whittaker associ\'ees
aux groupes de Chevalley}, Bull. Soc. Math. France, (1967),
{\bf 95}, 243-309.

\bibitem{Sch}G.Schiffmann, {\it Int\'egrales d'entrelacement et
fonctions de Whittaker}, Bull.Soc.Math. France, (1971),
{\bf 99}, 3-72.

\bibitem{Ha}M.Hashizume {\it Whittaker models for real
reductive groups},
J.Math.Soc. Japan, (1979), {\bf 5}, 394-401.\\
{\it Whittaker functions on semisimple Lie groups}, Hiroshima
Math.J., (1982), {\bf 12}, 259-293.

\bibitem{Skl}E.Sklyanin, {\it The quantum Toda chain}, Lect.Notes
in Phys., (1985), {\bf 226}, 196-233.

\bibitem{Konst}B.Kostant, {\it Quantization and representation
theory} In:
Representation theory of Lie Groups. Proc.SRC/LMS research Symp.,
Oxford 1977, London Math. Soc. Lecture Notes, {\it 34},287-316.

\bibitem{St}M.Semenov-Tian-Shansky. {\it Quantum Toda lattices.
Spectral theory and scattering.} Preprint LOMI R-3-84.
Leningrad, 1984. 64p. \\
M.Semenov-Tian-Shansky, {\it Quantization of Open Toda Lattices.}
Encyclopaedia of Mathematical Sciences, vol. 16. Dynamical
Systems VII. Ch. 3. Springer Verlag, 1994, pp.226-259.



\end{thebibliography}
\end{document}